\documentclass[10pt]{article} 
\usepackage{graphicx,graphics,floatflt,amssymb,epsf,psfig,rotate} 
\usepackage{color}

\def\ltap{\raisebox{-.4ex}{\rlap{$\sim$}} \raisebox{.4ex}{$<$}}

\begin{document} 
\begin{flushright} 
SINP/TNP/2008/13,  
HRI-P-08-06-002, 
CU-Physics/09-2008,
LPT-ORSAY 08-66
\end{flushright} 
 
\vskip 30pt 
 
\begin{center}
  {\Large \bf Radiative correction to the lightest neutral Higgs mass \\
   in warped supersymmetry} \\
  \vspace*{1cm} \renewcommand{\thefootnote}{\fnsymbol{footnote}} { {\sf Gautam
      Bhattacharyya${}^1$}, {\sf Swarup Kumar Majee${}^{2,3}$}, {\sf Tirtha
      Sankar Ray${}^{1}$}
  } \\
  \vspace{10pt} {\small ${}^{1)}$ {\em Saha Institute of Nuclear Physics,
      1/AF Bidhan Nagar, Kolkata 700064, India} \\
    ${}^{2)}$ {\em Harish-Chandra Research Institute,
      Chhatnag Road, Jhunsi, Allahabad  211019, India} \\
    ${}^{3)}$ {\em Department of Physics, University of Calcutta, 92
      A.P.C. Road, Kolkata 700009, India}}
 
\normalsize 
\end{center} 

\begin{abstract}  
  We compute radiative correction to the lightest neutral Higgs mass ($m_h$)
  induced by the Kaluza-Klein (KK) towers of fermions and sfermions in a
  minimal supersymmetric scenario embeded in a 5-dimensional warped space. The
  Higgs is confined to the TeV brane. The KK spectra of matter supermultiplets
  is tied to the explanation of the fermion mass hierarchy problem. We
  demonstrate that for a reasonable choice of extra-dimensional parameters,
  the KK-induced radiative correction can enhance the upper limit on $m_h$ by
  as much as 100 GeV beyond the 4d limit of 135 GeV.

\vskip 5pt \noindent 
\texttt{PACS Nos:~ 12.60.Jv, 11.10.Kk } \\ 
\texttt{Key Words:~~Supersymmetry, Warped Extra Dimension, Higgs mass}
\end{abstract}

\renewcommand{\thesection}{\Roman{section}} 
\setcounter{footnote}{0} 
\renewcommand{\thefootnote}{\arabic{footnote}} 
 
{\bf Introduction}:~ Minimal supersymmetric standard model (MSSM) with
superparticles in the 1 TeV range, primarily for its ability to settle the
gauge hierarchy problem and for providing a cold dark matter candidate, has
emerged as a leading candidate of physics beyond the standard model (SM). A
key prediction of MSSM is the existence of a light Higgs ($m_h <$ 135 GeV). If
such a light scalar exists, the CERN Large Hadron Collider (LHC) will find it
hard to miss. Moreover, if a quantum picture for all interactions including
gravity has to be weaved, we have to rely on string theory, which invariably
includes supersymmetry (SUSY). Since string theory is fundamentally a higher
dimensional theory, a reanalysis of the four-dimensional (4d) MSSM Higgs
spectra by embedding the theory in an extra-dimensional set-up is a worthwhile
phenomenological exercise.

Randall-Sundrum (RS) type models \cite{Randall:1999ee} with a warped
space-time geometry, where the bulk is a slice of Anti-deSitter (AdS) space in
which the SM/MSSM particles can also penetrate
\cite{bulksm,Gherghetta:2000qt}, lead to important phenomenological
consequences: (i) gauge hierarchy problem is solved thanks to the warp factor,
(ii) mass hierarchy of fermions can be explained by their relative
localizations in the bulk \cite{Huber:2000ie}, (iii) the smallness of neutrino
masses can be explained \cite{Grossman:1999ra}, (iv) gauge couplings unify if
the warped space is supersymmetric \cite{Dienes:1999sz}, (v) SUSY breaking can
be realized with a geometrical interpretation \cite{Gherghetta:2000kr}, (vi)
light Kaluza-Klein (KK) gauge boson and fermion states can be captured at the
LHC, and some other specific signals, like top flavor-violating decays, can be
detected as well \cite{rslhc}.

Since Higgs is the {\em most-wanted} entity at the LHC, our intention in this
paper is to calculate how the upper limit on the lightest supersymmetric
neutral Higgs mass changes in the warped extra-dimensional backdrop due to
radiative corrections induced by the KK towers of fermions and
sfermions. Before we perch on extra-dimensional details, we mention that even
within the 4d set-up the Higgs mass receives additional contribution, beyond
the MSSM limit of 135 GeV, in the next-to-minimal MSSM \cite{Drees:1988fc} and
in the left-right MSSM \cite{Zhang:2008jm}, to the tune of a few tens of a GeV
in cach case.

{\bf 5d warped MSSM}:~ The fifth dimension $y$ is compactified on an $S^1/Z_2$
orbifold of radius $R$. Two 3-branes are located at the orbifold fixed points
at $y = (0,\pi R)$. The space-time between the two branes is a slice of
AdS$_5$ geometry. The 5d metric is given by,
\begin{equation}
\label{metric}
        ds^2=e^{-2\sigma}\eta_{\mu\nu}dx^\mu dx^\nu+dy^2 ,~~~{\rm where}~~ 
\sigma=k|y|~.
\end{equation}
Above, $1/k$ is the AdS curvature radius and $\eta_{\mu\nu}={\rm
  diag}(-1,1,1,1)$.  The 3-brane at $y=0$ is called the Planck brane as the
natural mass scale associated with it is $M_P$, while the 3-brane at $y=\pi R$
with an effective mass scale $M_P e^{-\pi kR}$ can be called a TeV brane
provided $kR\simeq 12$.

We consider a supersymmetric scenario in which not only gravity but all
particles can, in principle, access the AdS bulk \cite{Gherghetta:2000qt}.
There are quite a few boons of supersymmetrizing an ${\rm AdS_5}$ slice,
e.g. a possible connection to string theory can be established, new avenues
for SUSY breaking can be explored, etc.  We concentrate on the structure of
matter hypermultiplets and the nature of Yukawa interactions for their special
relevance in the computation of radiative corrections to the Higgs mass.

The hypermultiplet $\Phi=(\phi^i,\Psi)$ contains two complex scalars $\phi^i$
($i=1,2$) and a Dirac fermion $\Psi$.  The action can be written as ($g \equiv
{\rm det} (g_{MN})$) \cite{Gherghetta:2000qt}
\begin{equation}
\label{kinhyper}
   S_5=-\int d^4x\int dy\sqrt{-g}\, \Bigg[
     \left|\partial_M \phi^i \right|^2+i\bar{\Psi}\gamma^MD_M\Psi
     +m^2_{\phi^i}|\phi^i|^2+im_\Psi\bar{\Psi}\Psi\Bigg]\, .   
\end{equation}
Invariance under supersymmetric transformation yields \cite{Gherghetta:2000qt}
\begin{eqnarray}
\label{h:susycon}
       m^2_{\phi^{1,2}}&=&(c^2\pm c-\frac{15}{4})k^2
       +\left(\frac{3}{2}\mp c\right)
          \frac{d^2\sigma}{dy^2} \, , ~~ {\rm and} ~~
       m_\Psi = c\frac{d\sigma}{dy} \, ,
\end{eqnarray}
where $c$ is some arbitrary real number. A generic 5d field can be decomposed
as \cite{Gherghetta:2000qt}
\begin{equation}
\label{Kaluza-Klein}
        \Phi(x^\mu,y)={1\over\sqrt{2\pi R}}\sum_{n=0}^\infty 
\Phi^{(n)}(x^\mu)f_n(y) , ~~{\rm where}~~
 f_n(y)=\frac{e^{s\sigma/2}}{N_n}\left[J_\alpha(\frac{m_n}{k}e^{\sigma})
     +b_{\alpha}(m_n)\, Y_\alpha(\frac{m_n}{k}e^{\sigma})\right]\, ,
\end{equation}
with $s = (4,1,2)$ for $\Phi=\{\phi,e^{-2\sigma}\Psi_{L,R},A_\mu\}$.  For the
detailed formulae of $b_{\alpha}$ and the normalization $N_n$ in terms of the
Bessel functions $J_\alpha$ and $Y_\alpha$ (where $\alpha$ is an index for 5d
mass and is a function of $c$), see Ref.~\cite{Gherghetta:2000qt}.
Regardless of $Z_2$ even or odd nature of KK modes, one obtains an approximate
expression of the KK mass:
\begin{equation}
\label{kkhypmassapp}
     m^{(n)}\simeq (n+\frac{c}{2}-\frac{1}{2}) \pi k e^{-\pi k R}~,~{\rm where}~
     n=1,2,..
\end{equation}
Two points are worth noting: (i) Even though 5d $N=1$ SUSY is equivalent to
$N=2$ in 4d, the massless sector of the hypermultiplet (from $\phi^1$ and
$\Psi_L$, in our convention) forms an $N=1$ chiral supermultiplet, and (ii) in
the conformal limit ($c=1/2$), the massless modes are not localized in the
bulk and couple to the two branes with equal strength. For $c<1/2$, the zero
modes are confined towards the TeV brane, while for $c>1/2$, the zero modes
are localized closer to the Planck brane.

We now come to the Yukawa interaction. First, we assume that the Higgs boson
is localized at the TeV brane, i.e. $H(x,y) = H(x) \delta(y-\pi R)$ [this
immediately solves the $\mu$ problem, as $\mu \sim \cal{O}$ (TeV)].
Considering that for each fermion flavor $i$, there are two 5d Dirac fermions
$\Psi_{iL}(x,y)$ and $\Psi_{iR}(x,y)$, one can write the Yukawa action as
($H(x) \equiv H_u(x) ~{\rm or}~ H_d(x)$)
\begin{equation}
\label{5dimenyc}
S_y=\int d^4x \int dy\, \sqrt{-g}\,\,\lambda_{ij(5d)} H(x) 
   \Big( \bar\Psi_{iL}(x,y)\Psi_{jR}(x,y) + {\rm h.c.} \Big)
\delta(y-\pi R) \, .
\end{equation}
Recall that each 5d fermion field has a bulk mass term, characterized by
$c_{iL}$ or $c_{iR}$. For simplicity, we assume that $c_i \equiv c_{iL}=
c_{iR}$. We now expand the 5d fermion fields in zero modes and higher KK
modes and obtain the corresponding 4d Yukawa couplings. For simplicity, we
consider only the diagonal couplings, i.e. ignore quark mixings as their
numerical effects are negligible for our calculation. The Yukawa couplings of
the zero mode fermions are given by
\begin{equation}
\label{yc}
    \lambda_{i}= \lambda_{i(5d)}k (1/2-c_i)
\left(1- e^{(2c_i-1)\pi kR} \right)^{-1} \, .
\end{equation}
Now we assume $\lambda_{i(5d)} k \sim 1$, and trade the zero mode fermion
masses in favor of the corresponding $c_i$ (see Table 1). This is how the
fermion mass hierarchy problem is addressed. We note here that the choice of
$c_i > 1/2$ for the first two families helps evade tight constraints ($m^{(1)}
>$ a few TeV) from FCNC processes \cite{bulksm}. For the third generation,
FCNC constraints are not so stringent any way.
\begin{center}
\begin{table}
\centering
\begin{tabular}{|c|c|c|c|c|c|c|c|c|c|} \hline
$f_i$&$e$&$\mu$&$\tau$&$u$&$d$&$c$&$s$&$t$&$b$\\  \hline
$c_i$&0.61&0.52&0.40&0.62&0.57&0.52&0.52&-0.50&0.26\\  \hline
$m^{(1)}_i$&1598&1508&1388&1609&1562&1504&1510&500&1249\\  \hline
\end{tabular}
\caption{\small \sf 
  The $c_i$ parameters and $m^{(1)}_i$ (in GeV) for different 
  flavors are shown for $kR=11.93$ and $\tan\beta=
{\langle H_u^0\rangle}/{\langle H_d^0\rangle} =
10$. For this choice, 
  the mass gap between the consecutive KK states is $m^{(n+1)} - m^{(n)} 
  = 1987$ GeV, irrespective of $c_i$. The corresponding $n=1$ KK mass for
  gauge boson is $1490$ GeV.}
\label{table}
\end{table}
\end{center}
\vskip -40pt We now turn our attention to the Yukawa couplings of KK
fermions. We {\em assume} KK number conservation at the tree level Higgs
coupling with the KK fermions\footnote{Although, unlike in UED, KK-parity is
  not automatic in the warped scenario, it is still possible to implement it
  in a slice of AdS$_5$ \cite{Agashe:2007jb}. We assume this parity for
  simplicity of our analytic computation. This also helps in evading some FCNC
  constraints.}. Inserting the expansion in Eq.~(\ref{Kaluza-Klein}) in
Eq.~(\ref{5dimenyc}), we derive (again, considering only diagonal couplings):
\begin{eqnarray}
\label{kkyukawa-full}
  {\lambda}_{i}^{(n)}& = &
  {\lambda}_{i(5d)} \pi m^{(n)} e^{\pi kR} {\left[
      J_{\alpha}\Big(\frac{m^{(n)}}{k}e^{\pi
        kR}\Big)+b_{\alpha}(m^{(n)})Y_{\alpha}\Big(\frac{m^{(n)}}{k}e^{\pi kR}\Big)
    \right] }^2 \, .
\end{eqnarray}
A set of (reasonable) approximations $m_n \ll k \sim M_P$, $kR \gg 1$ and
${\lambda}_{i(5d)}k \sim 1$ simplifies the above as, 
\begin{equation} 
  \label{kkyukawa}
  {\lambda}_{i}^{(n)} \sim {\cos}^2 \Big(\left[ n-\frac{3}{4} \mp \frac{1}{4}
  \right] \pi \Big) \, ,
\end{equation}
where $\mp$ correspond to $Z_2$ even/odd KK modes. We draw two important
conclusions from Eq.~(\ref{kkyukawa}): (i) all KK Yukawa couplings, regardless
of their flavors (i.e. $c_i$ values) and KK numbers, are roughly equal being
close to unity (more precisely, ${\lambda}_{i(5d)}k$), and (ii) the KK Yukawa
couplings of $Z_2$ odd modes are vanishing (since the Higgs is brane-bound).

{\bf Radiative correction to the Higgs mass}:~ We recall that in 4d MSSM the
lightest neutral Higgs mass at the tree level is lighter than $M_Z$, the
$Z$-boson mass. More specifically, $m_h \leq {\rm min}~(m_A, M_Z) |\cos
2\beta| \leq {\rm min}~(m_A, M_Z)$, where $M_A$ is the pseudo-scalar Higgs
mass.  The radiative correction to $m_h$, dominated by the top quark Yukawa
coupling and the masses of the stop squarks, is given by $\Delta m_h^2 \simeq
(3 m_t^4/2\pi^2 v^2) \ln (m_{\tilde{t}}^2/m_t^2)$. Here, 
$v^2 = {\langle H_u^0\rangle}^2 + {\langle H_d^0\rangle}^2
  = 1/(\sqrt{2} G_F) $ and $m_{\tilde{t}} =
\sqrt{m_{\tilde{t}_1} m_{\tilde{t}_2}}$ is an average stop mass \cite{dj}.

Here, our goal is to compute $\Delta m_h^2$ induced by the KK fermions. We
shall follow the effective potential technique to implement this correction.
We shall ignore the contributions from gauge sector as they are not
numerically significant. Dominant effects arise from matter sector.  The full
one-loop effective potential is given by $V_1(Q) = V_0(Q) + \Delta V_1(Q)$
where, in terms of the field dependent masses $M(H)$, $\Delta V_1(Q) =
{(64\pi^2)}^{-1}{\rm Str} M^4(H) \left \{ \ln(M^2(H)/Q^2) - 3/2 \right \}$.
The scale dependence of $\Delta V_1(Q)$ cancels against that of $V_0(Q)$
making $V_1(Q)$ scale independent up to higher loop orders.  The supertrace is
defined through ${\rm Str} f(m^2) = \sum_i (-1)^{2J_i} (2J_i+1) f(m_i^2)$,
where the sum is taken over all members of a supermultiplet.  The contribution
from the chiral multiplet labeled by the fermion $f$ is given by (with $N_c$ as
the color factor)
\begin{equation}
\label{epatop}
\Delta V_{f} = {N_c \over {32\pi^2}}
\left\{ m_{\tilde{f_1}}^4 \left(\ln{m_{\tilde
        {f_1}}^2\over{Q^2}}-{3\over2}\right)
  + m_{\tilde {f_2}}^4 \left(\ln{m_{\tilde
        {f_2}}^2\over{Q^2}}-{3\over2}\right)
  -2 m_{f}^4 \left(\ln{m_{f}^2\over{Q^2}}-{3\over2}\right)\right\} \, . 
\end{equation}
For illustration, we take only the up-quark chiral multiplet.  The field
dependent KK quark masses are,
\begin{equation} 
\label{mthmbh}
\left(m^{(n)}_{u}\right)^2(H) = \left(\lambda^{(n)}_{u}\right)^2 |H_u^0|^2 
+ \left(m^{(n)}\right)^2 ~,  
\end{equation}
while the KK mass square matrix of the up-squark is obtained as,   
\begin{eqnarray}
&& {M_{\tilde {{u}}}}^{(n)2} (H) =
\left(\begin{array}{cc} 
m_Q^2 + |H^0_{u}|^2 \left(\lambda^{(n)}\right)^2 
& {\lambda^{(n)}} (A_{{u}} H^0_{u}+\mu{H^0_d}^*)\\  
{\lambda^{(n)}} (A_{{u}}{H^0_u}^* +\mu H^0_d) & 
m_U^2 +|H^0_u|^2 \left(\lambda^{(n)}\right)^2  
\end{array}\right) +  \left(\begin{array}{cc} 
 \left(m^{(n)}\right)^2 & 0\\  
0 &  \left(m^{(n)}\right)^2   
\end{array}\right) \, . 
\label{tsquark}
\end{eqnarray} 
We treat the soft SUSY breaking parameters ($m_{Q,U}^2$, $A_u$), $\mu$ and
$\tan\beta$ as phenomenological inputs, assuming that the soft terms add in
{\em quadrature} (see Eq.~(\ref{tsquark}))\footnote{SUSY breaking by twisted
  boundary conditions on fermions leads to a {\em linear} splitting ($0.5 \pi
  k~ e^{-\pi kR}$) \cite{Gherghetta:2000kr}. The numerical effects on $\Delta
  m_h^2$ would be covered within the parameter space we scanned (see figures
  later).}. The radiative contribution to the (zero mode) CP-even Higgs
mass-square matrix from the $n$th KK quark/squark mode is given by,
\begin{eqnarray}
\label{msqevencor}
\left. \Delta{\cal{M}}^2\right|_{(n)ij} =
\left(N_c/{4\pi^2 v^2}\right) \left(\Delta^u_{ij} \right)^n \, ,
\end{eqnarray}
where, putting $\lambda^{(n)} = 1$ following Eq.~(\ref{kkyukawa}), and
denoting $g(a,b) \equiv 2-\{(a+b)/(a-b)\}\ln(a/b)$,
\begin{eqnarray}
\label{deltaup}
(\Delta_{11}^u)^n &=& {v_u^4\over{{4}{\rm sin}^2\beta}}\left(\mu (A_u+\mu
{\rm cot}\beta)\over{m_{\tilde u_1^n}^2 - m_{\tilde
u_2^n}^2}\right)^2g(m_{\tilde u_1^n}^2,m_{\tilde u_2^n}^2) ,\nonumber \\
(\Delta_{12}^u)^n &=& {v_u^4\over{{4}\sin^2\beta}}{\mu (A_u+\mu 
\cot\beta)\over{m_{\tilde u_1^n}^2 - m_{\tilde u_2^n}^2}}\left[\ln
{{m_{\tilde u_1^n}^2}\over{m_{\tilde u_2^n}^2}}+{A_u(A_u+\mu \cot
\beta)\over{m_{\tilde u_1^n}^2 - m_{\tilde u_2^n}^2}}g(m_{\tilde
u_1^n}^2,m_{\tilde u_2^n}^2)\right] , \\
(\Delta_{22}^u)^n& =& {v_u^4\over{{4}{\rm sin}^2\beta}}\left[{
\ln}{{m_{\tilde u_1^n}^2}{m_{\tilde u_2^n}^2}\over{m_{u^n}^4}} +
{2A_u(A_u+\mu \cot\beta)\over{m_{\tilde u_1^n}^2 -m_{\tilde
u_2^n}^2}} \ln {{m_{\tilde u_1^n}^2}\over{m_{\tilde
u_2^n}^2}} + \left(A_u(A_u+\mu
\cot\beta)\over{m_{\tilde u_1^n}^2 - m_{\tilde
u_2^n}^2}\right)^2g(m_{\tilde u_1^n}^2,m_{\tilde u_2^n}^2) \right] 
\nonumber \, .
\end{eqnarray}
The expressions for $(\Delta_{11}^d)^n$ can be written {\em mutatis
  mutandis}. 

A comparison with what happens in flat space supersymmetric Universal Extra
Dimension (UED) \cite{Bhattacharyya:2007te} is now in order. In UED, the KK
states are equispaced (due to space-time flatness), and the KK Yukawa
couplings are proportional to the corresponding zero mode masses. In the
warped scenario, the KK states have a sparse spectrum following the zeros of
the Bessel function, and the KK Yukawa couplings are, to a good approximation,
independent of the flavor indices and are all close to unity for a reasonable
choice of extra-dimensional parameters. In our warped SUSY scenario, for low
and moderate $\tan\beta (\sim 10)$, only $m_t^{(1)} \sim 500$ GeV is light,
which is a consequence of appropriately choosing the $c_i$ parameters for
correctly reproducing zero mode fermion masses. For large $\tan\beta(\sim
40)$, although $m_b^{(1)}$ can be as light as 500 GeV, the prefactor $v_d^4$
that appears in $(\Delta_{ij}^d)^n$ (not shown explicitly) suppresses the
radiative contribution to the Higgs mass induced by the down-type chiral
multiplets. So in the warped case, only $u^{(1)}, c^{(1)}$ and especially
$t^{(1)}$ multiplets contribute to $\Delta m_h^2$ in a numerically significant
way. The contributions from higher KK states are negligible.  This is in sharp
contrast with the SUSY UED scenario where the first {\em few} $t^{(n)}$ (and
{\em not} $u^{(n)}$ or $c^{(n)}$) chiral multiplets provide sizable
contribution to $\Delta m_h^2$. The net numerical effects in the two cases are
comparable. Recall that in UED, unlike in the warped case, the KK spectra are
not linked to fermion mass hierarchy.

\begin{figure*}
\begin{minipage}[t]{0.42\textwidth}
\begin{center}
\includegraphics[width=0.7\textwidth,angle=270,keepaspectratio]
{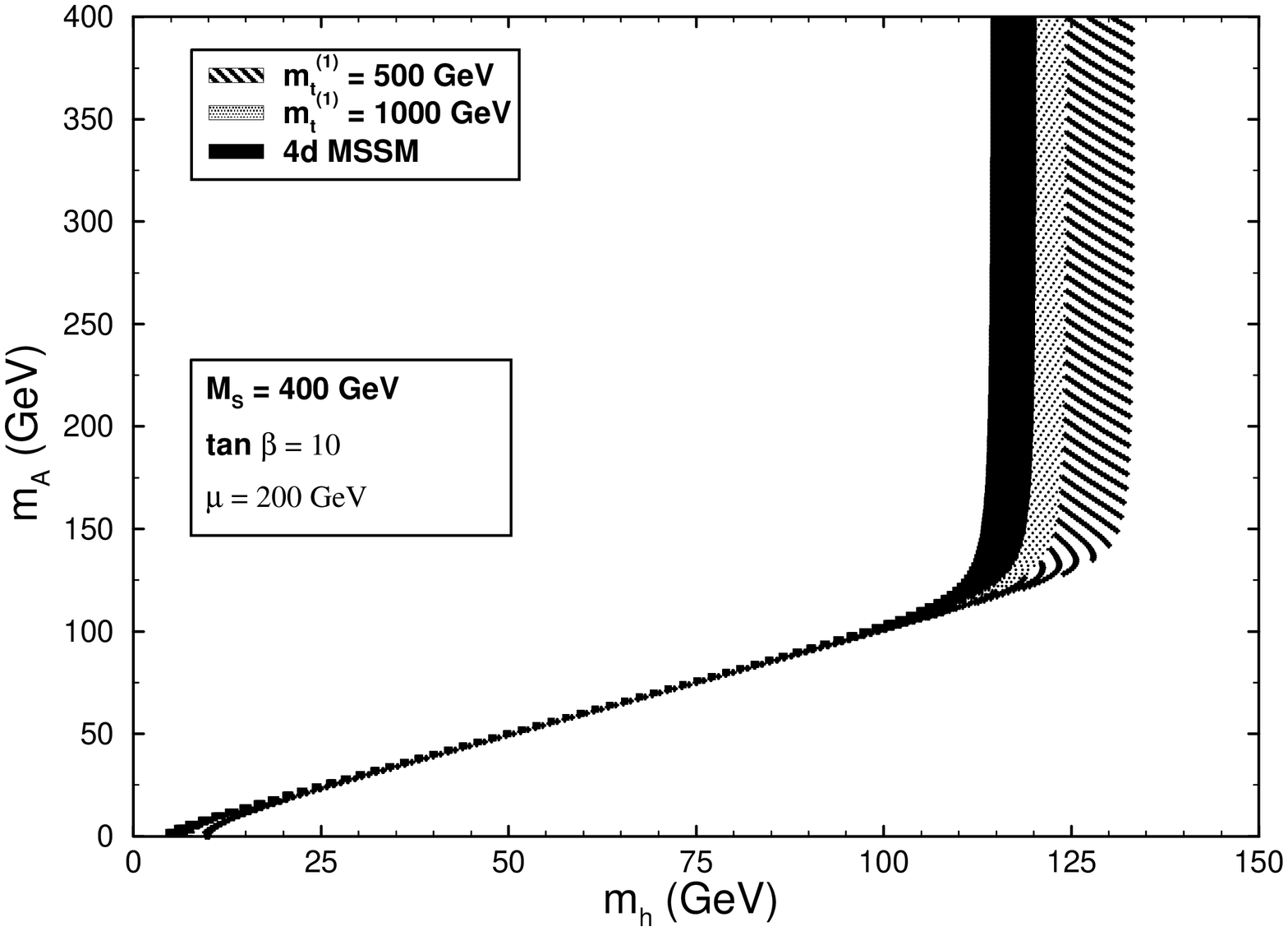}
\caption[]{\sf \small A comparison between 4d MSSM and 5d warped MSSM in the
  $m_h$-$M_A$ plane. The band width corresponds to the variation of $A_u$ and
  $A_d$ in the range $(0.8 - 1.2)M_S$.}
\end{center}
\end{minipage}
\hspace{7mm}
\begin{minipage}[t]{0.42\textwidth}
\begin{center}
\includegraphics[width=0.7\textwidth,angle=270,keepaspectratio]
{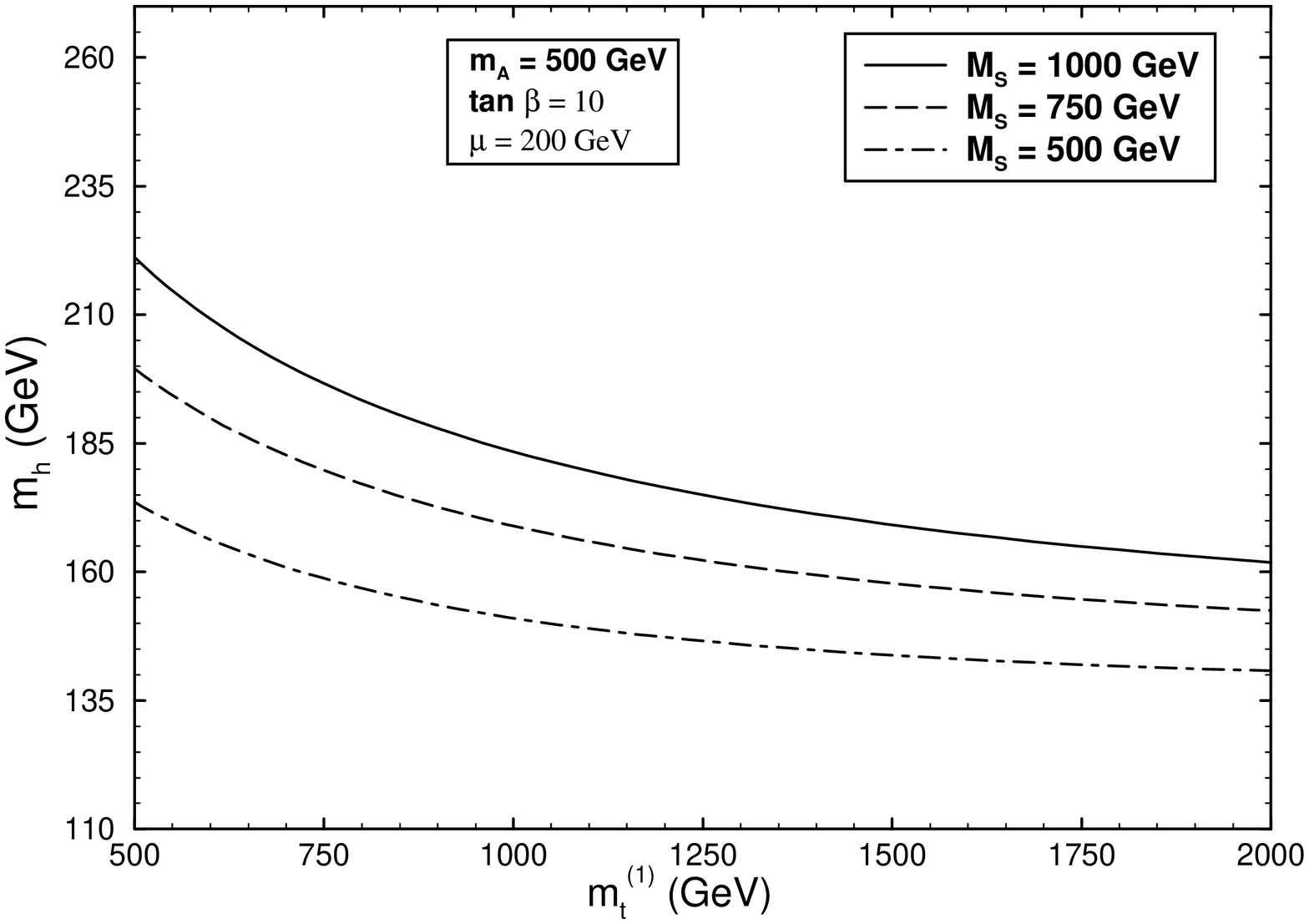}
\caption[]{\sf \small The variation of $m_h^{\rm max}$ with $m_t^{(1)}$ for
  different choices of $M_S$. We have used $A_u = A_d = \sqrt{6} M_S$ to
  maximise the radiative effect.}
\end{center}
\end{minipage}
\end{figure*}
{\bf Numerical Results}:~ For simplicity, we have assumed a common soft mass
$M_S\equiv m_Q=m_U=m_D$. The trilinear couplings $A_u$ and $A_d$ have been
varied in the range $[0.8-1.2]M_S$, which resulted in bands in the figures.
In Fig.~1, we have plotted our results in the $m_h$-$m_A$ plane.  We have used
$m_t^{(1)}$, the lightest KK top quark mass ({\em modulo} the zero mode mass),
to represent the extra-dimensional effect. 
In Fig.~1, we have displayed the effects for two choices of $m_t^{(1)}$,
namely, 500 and 1000 GeV, and for a moderate $\tan\beta=10$. In Fig.~2, we
have demonstrated that $m_h$ indeed falls with increasing $m_t^{(1)}$,
eventually attaining its 4d value. In this plot, we have set
$A_u=A_d=\sqrt{6}M_S$, which maximises not only the 4d MSSM radiative
correction but also the KK-induced one, which is why we have used the symbol
$m_h^{\rm max}$. The three lines correspond to three different choices of $M_S
=$ 500, 750 and 1000 GeV. All in all, $m_h$ increases by a few to several tens
of a GeV, depending on the choice of soft SUSY breaking parameters, the
radiative contribution coming primarily from all up-type multiplets. Beyond
$m_t^{(1)} = 2$ TeV (which corresponds to an average value of $\sim$ 6 TeV for
other first level KK fermions and gauge bosons), the KK effects almost vanish
and the three falling curves become flat lines approaching their 4d limits (of
course, without ever meeting as the three $M_S$ values are different).

{\bf Conclusions}:~ We have calculated one-loop correction to the lightest
neutral Higgs boson mass in a generic MSSM embeded in a slice of AdS$_5$. For
a reasonable choice of warped space parameters, the 4d upper limit of 135 GeV
could be relaxed by as much as $\sim$ (50-100) GeV depending on $M_S$. A few
other closely related highlights are the following: (i) matter KK spectra are
controlled by the $c_i$ parameters, which, in turn, are determined by the zero
mode fermion masses; (ii) all KK Yukawa couplings are close to unity to a very
good approximation; (iii) only $m_t^{(1)}$ could be light ($\sim$ 500 GeV) for
a moderate $\tan\beta$ ($m_b^{(1)}$ can also be light too for large $\tan\beta
\sim 40$) - this feature can be used to discriminate this scenario from the
others producing similar shift to $m_h$; (iv) small values of $\tan\beta
(\ltap~3)$, which are otherwise disfavored in 4d MSSM due to nonobservation of
Higgs up to 114.5 GeV \cite{lep}, are now resurrected thanks to an additional
KK-induced radiative corrections. Admittedly, the stability of the proton
would require further care \cite{Gherghetta:2000qt}. Besides, the warped
models with fermions in the bulk, in general, pass the electroweak precision
tests (EWPT) with some difficulty \cite{Hewett:2002fe}, unless the KK mass is
raised to tens of a TeV. To suppress excessive contribution to $T$ (or $\Delta
\rho$), gauge symmetry in the bulk is enhanced to ${\rm SU(2)_L \times SU(2)_R
  \times U(1)_{B-L}}$ \cite{Agashe:2003zs}, while to keep the contributions to
$Zb_L\bar b_L$ vertex and other loop corrections under control, a further
discrete $L \leftrightarrow R$ symmetry has been employed
\cite{Carena:2007ua}. This allows us to consider $m_t^{(1)}$ as light as 500
GeV (the lightest KK gauge boson is then 1.5 TeV). Furthermore, $\tan\beta$
can be tuned to reduce the contribution to $T$.  Since our primary intention
here has been to develop a simple analytic framework to (for the first time)
compute KK-induced radiative corrections to $m_h$ in a supersymmetric warped
space, we pared the scenario down to its bare minimum. The further details
necessary to overcome the above constraints are unlikely to alter the
essential qualitative and quantitative features we explored here.

Finally, we note that in a general class of such models, KK-parity violating
gauge interactions would induce the lightest KK particle decay into zero mode
SM particles well before nucleosynthesis sets in, thus without disturbing any
cosmological constraints. The dark matter candidate (which could very well be
the lightest zero mode neutralino, provided $R$-parity is conserved) would
still be decided on the basis of the 4d spectra controlled by our choice of
soft masses. Moreover, the lowest KK excitations of the SM particles could be
heavier or lighter than the zero mode superparticles, again depending on our
choice of the zero mode supersymmetry breaking soft parameters.  A detailed
phenomenological study preparing a catalogue of all such possibilities is
beyond the goal of this paper.

\noindent{ \bf Acknowledgements:}
We thank E.~Dudas and A.~Raychaudhuri for valuable comments. GB acknowledges
hospitality at LPT-UMR 8627, Orsay, and a CNRS fellowship during a part of
this work, and a partial support through the project No.~2007/37/9/BRNS of
BRNS (DAE), India. TSR acknowledges the S.P. Mukherjee fellowship of CSIR,
India.

\end{document}